\RequirePackage{lineno}
\documentclass[%
12pt,%
reprint,
showpacs,
amsmath,amssymb,aps,
]{revtex4-1}

\usepackage{amssymb,amsmath,amsfonts} 
\usepackage{graphicx,color,float,bm}

\usepackage{caption,subcaption}

\let\oldalign\align
\let\endoldalign\endalign
\renewenvironment{align}{%
  \linenomath
  \oldalign
}{%
  \endoldalign
  \endlinenomath
}

\def\eeq{\relax}
\def\beq#1#2\eeq{\begin{equation}\label{#1}#2\end{equation}}
\def\bal#1#2\eal{\begin{align}\label{#1}#2\end{align}}
\def\bse#1#2\ese{\begin{subequations}\label{#1}#2\end{subequations}}
\def\ba{\begin{aligned}}   \def\ea{\end{aligned}}

\def\XXint#1#2#3{{\setbox0=\hbox{$#1{#2#3}{\int}$}
\vcenter{\hbox{$#2#3$}}\kern-.5\wd0}}

\def\i{\operatorname{i}} 
\def\e{\operatorname{e}} 
\def\div{\operatorname{div}} 
\def\grad{\operatorname{grad}} 
\def\tr{\operatorname{tr}} 
\def\dd{\operatorname{d}} 
\def\Im{\operatorname{Im}}
\def\Sym{\operatorname{Sym}}
\def\Re{\operatorname{Re}}

\begin{document} 
\title{
\textcolor{blue}
{An inequality for longitudinal and transverse wave attenuation coefficients
 }  
}
\author{ Andrew N. Norris} 							
\email{norris@rutgers.edu}
\affiliation{Mechanical and Aerospace Engineering, Rutgers University, 98 Brett Road, Piscataway, NJ 08854}

\date{\today}
\pacs{43.20.Gp, 43.40.Dx, 43.35.Gk, 43.20.Tb}

%%%%%%%%%%%%%%%%%%%%%%%%%%%%%%%%%%%%%%%%%%%%%%%%%%%%%%%%%%%%%%%%%%%%%%%%%%%
%\def\singlespacing{\baselineskip=13pt}	\def\doublespacing{\baselineskip=18pt}
%\singlespacing%\doublespacing

%\pagestyle{myheadings}\markright{\currfilename \qquad    ~~~~~~\today}

\begin{abstract} %%%%%%%%%%%%%%%%
Total  absorption,  defined as the net flux of energy out of a bounded region averaged over one cycle for time harmonic motion, must be non-negative when there are no sources of energy within the region.  
This passivity condition  places constraints on the non-dimensional absorption coefficients of  longitudinal and transverse waves,   $\gamma_L$ and    $\gamma_T$, in  isotropic  linearly viscoelastic materials.    Typically,  $\gamma_L, \, \gamma_T$ are small, in which case the constraints imply that  coefficients of attenuation  per unit length,   $\alpha_L$,    $\alpha_T$,  must satisfy the inequality 
 ${\alpha_L}/{\alpha_T} \ge { 4c_{T}^3} / { 3c_{L}^3}$ where $c_L$, $c_T$ are the wave speeds.     This  inequality, which as far as the author is aware, has not been presented before, provides a relative bound on wave speed in terms of attenuation, or {\it vice versa}.  It also serves as a check on the consistency of  ultrasonic measurements from the literature, with most but not all of the data considered passing the positive absorption test. 

\end{abstract}   %%%%%%%%%%%%%%%%

\maketitle

\section{Introduction}\label{sec1}

%The motivation for this short note arose in c
When  an acoustic wave is incident on a passive obstacle, i.e. one with no active sources of energy present, the scattering process should not result in  more energy than that of the incident wave itself.  If the object is  viscoelastic then the total energy should  decrease by virtue of the passive absorbing properties of viscoelastic solids.  {The absorption of wave energy in solids has been considered from many points of view, ranging from  its thermodynamic and molecular origins \cite[Chs.\ 11-13]{Bhatia67}, to sub-wavelength scattering effects at the crystalline and granular scale  \cite[Ch.\ 9]{Schmerr2016}, to  physically  consistent mathematical models \cite[Chs.\ 3.E, 5.C-E]{AuldI} \cite[Ch.\ 2]{Carcione2001}.  
The interest here is in linearly viscoelastic materials, and the implications of positive absorption, also known as passivity \cite{Weaver1981}.
%Our objective here is to describe some consequences of the fact that absorption in linearly viscoelastic solids, as defined more precisely below, is always non-negative.  
In particular, we derive a new and useful relation between the attenuation coefficients for longitudinal and transverse waves isotropic solids.   
}

We begin in Section \ref{sec2} with a background review of absorption and viscoelasticity theory.  The main results are presented in  Section \ref{sec3}, where they are discussed in the context of published data on ultrasonic attenuation coefficients. 

\section{Background review}\label{sec2}

\subsection{Absorption}

{The energy lost in a passive target subject to an incident time harmonic acoustic  wave is defined by  the   time-averaged outward net flux  over the bounding  surface of  the object, 
%$P_\text{abs} \ge 0$. 
\bal{3}
P_\text{abs} &= -\int_S \langle { p{\bf v} } \rangle \cdot \dd {\bf s} 
\notag \\ & \ge 0.
\eal
Here,   $p$ denotes acoustic pressure,   ${\bf v}$ the particle velocity, $S$ is the enclosing surface,  
$\dd {\bf s}= {\bf n} \dd S$ is the surface element with the unit normal ${\bf n}$ outwards, and $\langle\cdot \rangle$ indicates the  average over a period. 
%uniform  acoustic medium of density $\rho_0$ and bulk modulus $\kappa_0$, with acoustic parameters  
%particle velocity ${\bf v}$ and acoustic pressure $p$. The linearized equation of motion  and the constitutive equation for acoustic pressure, 
%\bse{1}
%\bal{1a}
%\rho_0 \partial_t {\bf v} + \grad p &=0, 
%\\
 %\partial_t p + \kappa_0 \div {\bf v} &= 0,  
%\eal
%\ese
%together yield the energy equation
%\beq{2}
%\frac 12 \partial_t \big( \rho_0 v^2 + \kappa_0^{-1} p^2\big) + \div p{\bf v} = 0. 
%\eeq
%Both $p$ and ${\bf v}$ are periodic functions for  time harmonic motion of single frequency, so that averaging over the unit period implies that the averaged energy flux is divergence free.  That is, with  an  overline denoting  the average,
%\beq{2.1}
%\div \overline{ p{\bf v} } = 0. 
%\eeq
%Now consider a volume region $V$ which  may contain scatterers of any form. Our interest is in the  absorption  defined as the time-averaged outward net flux 
%\beq{3}
%P_\text{abs} = -\int_S \overline{ p{\bf v} } \cdot \dd {\bf s}, 
%\eeq
%where $S$ is the enclosing surface    and 
%$\dd {\bf s}= {\bf n} \dd S$ is the surface element, with the unit normal ${\bf n}$ outwards. 
%The region may contain  fluid and other obstacles.  
To be specific consider a single solid object of volume $V$.  Traction  continuity across $S$ implies 
$ - p {\bf n} = \boldsymbol{\sigma}{\bf n}$  where  $\boldsymbol{\sigma}$ is  the symmetric stress in the solid.  
}Normal velocity is also continuous, %therefore $-   { p{\bf v} }   \cdot \dd {\bf s} =  { \boldsymbol{\sigma}{\bf v} }  \cdot\dd {\bf s}$, 
%\beq{4}
%P_\text{abs} = \int_S \overline{ \boldsymbol{\sigma}{\bf v} } \cdot\dd {\bf s}, 
%\eeq
and the divergence theorem therefore implies the equivalent definition %and well known identity \cite{AuldI}
\beq{5}
P_\text{abs} = \int_V \div \langle{ \boldsymbol{\sigma}{\bf v} }\rangle \dd V   . 
\eeq

\subsection{Viscoelasticity}

%The author could not find much consideration of this problem in the  literature, with the notable exception  \cite{Srivastava2015}.   The approach in the latter is slightly different in that the final results are phrased in terms of the complex-valued compliance tensor, whereas here we find conditions for the stiffness tensor.  Since these are closely related inverse quantities, it is believed that the present findings are in agreement with those of \cite{Srivastava2015}.  
%However, here we consider more specifically standard formulations of linear viscoelasticity as used in acoustics and ultrasonics, and derive constraints on  commonly used  parameters for measuring wave damping, specifically  absorption and attenuation coefficients for longitudinal and transverse waves. 

 In order to accommodate  a viscoelastic constitutive relation  it is necessary to work with complex-valued quantities.  The real and imaginary parts of material properties, such as density $\rho$ and elastic stiffness ${\bf C}$, are denoted in standard fashion using single and double primes: 
\beq{-5}
\rho  =\rho '+ \i \rho '', \ \ 
{\bf C} = {\bf C}' + \i {\bf C} ''.
\eeq
{The density  is also  considered complex-valued as this represents a better alternative  to using viscoelastic moduli in certain materials encountered especially in geophysical acoustics, such as poroelastic continua where the complex $\rho$ includes Darcy-like flow effects on the overall inertia   \cite{JohnsonKoplikDashen1987,DupuyStovas2014}.}

Notwithstanding the danger of confusion, we now let the field variables $\boldsymbol{\sigma}$ and ${\bf v} $ (until now assumed to be real quantities) denote  complex-valued amplitudes with the time dependence $\e^{-\i\omega t}$, $\omega >0$,   understood.  The real physical quantities are  
$\Re \boldsymbol{\sigma}({\bf x})\e^{-\i\omega t} $ and $\Re {\bf v} ({\bf x})\e^{-\i\omega t} $.  With $^*$ denoting the complex-conjugate, \eqref{5} becomes 
\beq{6}
P_\text{abs} = \frac 12 \Re \int_V \div { \boldsymbol{\sigma}{\bf v}^* } \dd V.   
\eeq 
Using the equation of motion, 
\beq{7}
\div \boldsymbol{\sigma} +\i\omega \rho  {\bf v}  =0 .
\eeq	
the absorption can be expressed 
\beq{8}
P_\text{abs} = \frac {\omega}2  \int_V\rho '' |{\bf v}|^2 \dd V 
+\frac12  \Re \int_V  \tr ( \boldsymbol{\sigma} \grad {\bf v}^* ) \dd V   .
\eeq

Strain is the symmetric part of the displacement gradient
%\beq{9}
$\boldsymbol{\varepsilon} = \Sym {\bf U}$, %\ \ 
${\bf U} = \grad {\bf u}$, 
%\eeq
where ${\bf u} = (-\i\omega)^{-1} {\bf v}$ is the displacement.    

{The  viscoelastic linear constitutive relation  between stress and strain in its most general form 
\cite[Ch.\ 4]{Carcione2001} posits stress as a convolution of strain with a time dependent stiffness.  
The relation is then  linear in the frequency domain, 
\beq{11}
 \boldsymbol{\sigma} = {\bf C} \boldsymbol{\varepsilon}  \ \ \Leftrightarrow \ \ 
{\sigma}_{ij} = C_{ijkl} {\varepsilon}_{kl} 
\eeq
with  complex-valued stiffness ${\bf C}$ defined by the Fourier transform of the time-dependent moduli.  The latter are assumed to have the symmetries associated with a symmetric strain and a symmetric stress, implying 
}
$C_{ijkl} = C_{ijlk}$,  $C_{ijkl} =  C_{jikl}$. 
Hence, $\boldsymbol{\sigma} = {\bf C} {\bf U}$, and 
$ \Re  \tr ( \boldsymbol{\sigma} \grad {\bf v}^* )  =
-\omega \Im    \tr  {\bf U}^*{\bf C} {\bf U}$.
In purely elastic solids the stiffness  ${\bf C}$ is real-valued and  satisfies the usual symmetry in terms of the interchange  of the "major indices" associated with a reversible strain energy  function.  This property does not extend to viscoelasticity. However, it is expected in the quasistatic limit, and it is therefore reasonable to assume that it holds for the real part of ${\bf C}$, at least in some range of frequencies, but is not valid for the imaginary part of 
${\bf C}$. We therefore split the imaginary part into symmetric and anti-symmetric parts, 
\beq{132}
\begin{aligned}
  C_{ijkl}' &=  C_{klij}'; 
\ \    {\bf C}'' =  {\bf C}_S '' +  {\bf C}_A '', \\  
C_{Sijkl} '' &=  C_{Sklij} '', \  C_{Aijkl} '' =  -C_{Aklij} ''.
\end{aligned}
\eeq
The absorption can then be written 
\beq{14}
P_\text{abs} = \frac {\omega}2  \int_V \big( \rho'' |{\bf v}|^2 
- \tr {\bf U}^*{\bf C}_S'' {\bf U}\big) \dd V .  
\eeq
This is always non-negative if and only if $ \rho''$ is non-negative and ${\bf C}_S ''$ is negative semi-definite in the sense that 
%\beq{15}
 $\tr {\bf A}^*{\bf C}_S '' {\bf A} \le 0 $ for all ${\bf A}={\bf A}^T \ne 0$. 
%\eeq
In summary, 
\beq{17} 
P_\text{abs} \ge 0 \   \Leftrightarrow  \ \rho''\ge 0 \  \text{and} \ 
{\bf C}_S '' \  \text{is negative semi-definite} . 
\eeq
The absorption is identically zero if the density is purely real and ${\bf C}_S ''$ vanishes.  
The latter condition is equivalent to the requirement that ${\bf C}$ is Hermitian, i.e. 
$C_{ijkl} = C_{klij}^*$. 
Hence, 
\beq{171} 
P_\text{abs} = 0 \  \ \Leftrightarrow  \ \rho  \text{ is real and} \ 
{\bf C} \  \text{is Hermitian} . 
\eeq
Note that the present results do not rely upon the necessary consequences of causality on the analytic properties of the complex-valued moduli, a topic that has been addressed well elsewhere, e.g. \cite{Pritz1998}.

\subsubsection{Isotropic viscoelasticity}
The moduli have standard form with two complex-valued Lam\'e  moduli, $\lambda$ and $\mu$, %the bulk and shear modulus, %with  $\lambda= \lambda_0 + \i\lambda '$,  $\mu  = \mu_0 +\i\mu'$, 
\beq{18}
C_{ijkl} = \lambda \delta_{ij}\delta_{kl} + \mu( \delta_{ik}\delta_{jl} + \delta_{il}\delta_{jk}). 
\eeq
This stiffness satisfies \eqref{132} with ${\bf C}_A''=0$, i.e. ${\bf C}$ is not Hermitian, and    it is therefore expected that $P_\text{abs}$ will  be non-zero.  
The absorption becomes
\beq{19}
P_\text{abs} = \frac {\omega}2  \int_V \big( \rho'' |{\bf v}|^2 
- 
\kappa '' |\tr \boldsymbol{\varepsilon} |^2 -\mu '' \tr(  \boldsymbol{\varepsilon}_d  \boldsymbol{\varepsilon}_d^* )
\big) \dd V ,   
\eeq
where  $\kappa = \lambda +\frac 23 \mu$ is the bulk modulus and 
$\boldsymbol{\varepsilon}_d = \boldsymbol{\varepsilon} - {\bf I} \frac 13 \tr \boldsymbol{\varepsilon}$ is the deviatoric strain. 
Hence, 
\beq{20} 
P_\text{abs} \ge 0 \  \ \Leftrightarrow  \ \rho''\ge 0,  \  
\kappa '' \le 0 \ 
\text{and} \ 
\mu '' \le 0. 
\eeq
This places constraints on the imaginary parts of the elastic moduli. %, distinct from the  Hermitian constraint \eqref{17}.

%\subsection{Commonly used visocelastic models}
For instance, the   Kelvin-Voigt model assumes that the stress is of the form 
\beq{41}
\boldsymbol{\sigma} = ( \lambda_e + \lambda_v \partial_t) (\tr \boldsymbol{\varepsilon} ) {\bf I}
+ 2(  \mu_e + \mu_v \partial_t)  \boldsymbol{\varepsilon} 
\eeq
where $\lambda_e$, $\mu_e$ are the elastic moduli, $\lambda_v$, $\mu_v$ are generalized viscosities, all real quantities.  Hence, 
$\lambda '' = -\i \omega \lambda_v$,  $\mu '' =  -\i \omega \mu_v$, and the 
 constraints \eqref{20} are satisfied if $\lambda_v+ \frac 23 \mu_v >0$ and $\mu_v >0$.  

Positive absorption has implications for other elastic moduli. 
For any complex-valued elastic modulus $M= M'+\i M''$,  the loss factor \cite[p.\ 7]{readanddean} is defined as $d_M$$=$$ -M'' /M'$, so that $M = ( 1- \i d_M) M' $.   The constraints \eqref{20} imply that $d_\mu \ge 0$,  $d_\kappa \ge 0$. 
  The longitudinal modulus, Poisson's ratio, Young's modulus and area modulus
	\cite{Scott00} are $L = \lambda + 2\mu$, $\nu =  \lambda/[2( \lambda + \mu)]$, 
	$E = 2(1+\nu) \mu$ and $A = (1-\nu)^{-1}(1+\nu) \mu$,  respectively.  Their loss factors are, to leading order in $d_\mu  $ and  $d_\kappa $, 
\bse{90}
\bal{90a}
 d_A &= c_1 d_\mu  +  (1- c_1) d_\kappa ,
\\
d_L &=  (1-c_1 )d_\mu  +   c_1 d_\kappa ,
\label{90b}
\\
d_E &= c_2 d_\mu  +  (1- c_2) d_\kappa ,
\label{90c}
 %d_E &= \frac 23 (1+\nu ')d_\mu  + \frac 13 (1-2\nu ') d_\kappa ,
%\\
%d_L &= \frac 1{3(1-\nu ')} \big[ 2 (1-2\nu ')d_\mu  +  (1+\nu ')  d_\kappa \big], 
%\label{90b}
%\\
%d_A &= \frac 1{3(1-\nu ')} \big[   (1+\nu ')d_\mu  + 2 (1-2\nu ')  d_\kappa \big], 
%\label{90c}
\\
d_\nu &= (3 \nu ')^{-1} (1+\nu ')  (1-2\nu ')( d_\mu  - d_\kappa ),
\label{90d} 
\eal
\ese
where $c_1 = \frac 13 (1+\nu ')/(1-\nu ')$, $c_2 = \frac 23 (1+\nu ')$. 
%The identities \eqref{90b}, \eqref{90c} and \eqref{90d} correspond to eqs.\ (8.11), (8.14) and (8.9) of \cite{readanddean}, respectively. 
Positive definiteness of the elastic strain energy requires that  $-1< \nu ' < \frac 12$, and hence $0< c_1 < c_2 < 1$ and the loss factors  $d_L$, $d_E$ and $d_A$ are always non-negative 
with values between $ d_\mu$ and $ d_\kappa$.  The Poisson's ratio loss factor may  in principle be of either sign, although reported values, e.g. for rubber  \cite{Pritz1998}, are  positive indicating $ d_\mu  > d_\kappa $.

\section{Elastic wave damping }\label{sec3}

\subsection{Constraints on absorption coefficients }

The complex-valued longitudinal and transverse wavenumbers, $\tilde k_L$ and $\tilde k_T$,  are   
\beq{211}
\tilde k_L = \frac{\omega}{\tilde c_L}, \ 
\tilde k_T = \frac{\omega}{\tilde c_T}, \ 
%\text{where} 
\  \tilde c_L = \sqrt{\frac{\lambda+2\mu}\rho}, \ \tilde c_T = \sqrt{\frac{\mu}\rho}.
\eeq
A common method for characterizing viscoelasticity is via ultrasonic measurement
of the complex-valued wavenumbers.  Specifically, we assume that the viscoelastic moduli are defined  in terms of two   real-valued wave speeds $c_{L}$, $c_{T}$ and two  non-dimensional absorption coefficients
$\gamma_L$, $\gamma_T$: 
\beq{21}
\tilde k_M = k_M ( 1+ \i \gamma_M)  \ \ \text{with} \  k_M = \frac{\omega }{c_M}, \ \ M=L,\ T .
\eeq
According to this definition, $c_M = \big(\Re {\tilde c_M}^{-1} \big)^{-1}$, $M=L,\ T$.  
%Writing $\rho = \rho_e +i \rho'$,  $\kappa = \kappa_e +i \kappa '$,  $\mu = \mu_e +i \mu '$, then i
It follows from the imaginary parts of the identities 
${\tilde c_M}^2  = c_M^2/(1+ \i \gamma_M)^2$, $M=L,\ T$ that 
\bse{-21}
\bal{-21a}
%\mu_e \rho ' - \rho_e \mu' 
 \Im \frac{\mu}{\rho} &=  -\frac{2c_T^2 \gamma_T}{(1+\gamma_T^2)^2}, 
\\
%\kappa_e \rho ' - \rho_e \kappa '
 \Im \frac{\kappa}{\rho}   &=   -\frac{2c_L^2\gamma_L}{(1+\gamma_L^2)^2}
+   \frac{8c_T^2\gamma_T}{3(1+\gamma_T^2)^2}  .
\eal
\ese
The three constraints of \eqref{20} imply that the left members in \eqref{-21} are non-positive, and hence we obtain the main result of the paper: 
\beq{23}
\gamma_T \ge 0, 
\ \  
\gamma \ge 0; 
\ \ %\ \text{where } \ 
\gamma \equiv 
 \frac{\gamma_L}{(1+\gamma_L^2)^2} - \frac {4 {c_{T}}^2}{3 c_{L}^2}  \frac{\gamma_T}{(1+\gamma_T^2)^2} .
\eeq
The first is usually satisfied because both $\gamma_T $ and $\gamma_L $ are specifically taken as non-negative.  The condition for $\gamma $ places a constraint on $\gamma_T $ and $\gamma_L $ that depends upon  the ratio of the undamped wave speeds.    
The value of $\gamma$  for measurements on 
Polymethylmethacrylate (PMMA) and other polymers   are given  in Table \ref{table1}, 
all    satisfying the condition $\gamma >0$.

\begin{table}[h]
\begin{center}
\begin{tabular}{lcccccc} %{llllll}
\hline \hline
%\\
Material &  Source & $c_L$  &  $c_T$   & $\gamma_L$ & $\gamma_T$  & $\gamma$
\\
\hline 
PMMA &  \cite{Hartmann1972}& 2690 &1340 & 0.0035 & 0.0053 &0.0017
\\
Polyethylene & \cite{Hartmann1972}&2430 &950 & 0.0073 & 0.0220 & 0.0028
\\
Phenolic 
polymer & \cite{Hartmann1975}&
 2840 &1320 &0.0119& 0.0255   & 0.0045
\\
\hline \hline
\end{tabular}
\caption{An example of some absorption coefficients and the  associated value of  $\gamma $ from  eq.\ 
\eqref{23}.  The numerical values from \cite{Hartmann1972} are for measurements  at room temperature of  $\alpha \lambda = 2\pi \gamma$, the attenuation per wavelength in dB, and uses the relation \cite{Schuetz1977}  $\alpha \lambda $ (dB) $= 40\pi\gamma /\ln 10$.  
Those of  \cite{Hartmann1975} are based on measurements of the attenuation coefficients $\alpha_L$ and $\alpha_L$ (dB/cm) at 25$^\circ$C and frequency 1.8 MHz. 
Speeds are in m/s.}
\label{table1}
\end{center}
\end{table}

In practice the values of $\gamma_L$ and  $\gamma_T$ are small, so that \eqref{23}$_2$ $(\gamma \ge 0)$ can be safely replaced by 
\beq{231}
%\gamma_T \ge 0, \ \  
\frac{\gamma_L}{\gamma_T} \ge 
\frac { 4{c_{T}}^2} { 3c_{L}^2} \ \ \text{for } \gamma_L, \gamma_T \ll 1.
\eeq
Thus, the ratio of the wave absorption factors must satisfy a strict but simple inequality when the attenuation is small. Note that the parameter depends upon the real part of the Poisson's ratio, 
\beq{232}
 \frac { 4{c_{T}}^2} { 3c_{L}^2}= \frac{2(1-2\nu ')}{3(1-\nu')} .
\eeq

\subsection{Constraints on attenuation coefficients}

The amplitude of either wave type decays as  $e^{-k_M\gamma_M x}$, $M=L$ or $T$.  Attenuation as measured in dB/cm, for instance, defines  the logarithm of the amplitude, and is therefore equivalent to measurement of 
$\alpha_M \equiv k_M  \gamma_M$, $M=L$ or $T$, since 
$\alpha = k\gamma \, 20/ \ln 10$ where $k$ is wavenumber in cm$^{-1}$.  The multiplicative factor is irrelevant if we are only concerned with the ratio of the two attenuations.  In the small attenuation regime eq.\ \eqref{231} then implies 
\beq{25}
\frac{\alpha_L}{\alpha_T} \ge 
\frac { 4c_{T}^3}  { 3c_{L}^3} .
\eeq

\begin{table}[H]
\begin{center}
\vspace{.2in}
\begin{tabular}{ccccccc} %{llllll}
\hline \hline
Material &  $f$  &   $c_L$ &  $c_T$   & $\alpha_L$   & $\alpha_T$    &  Eq.\ \eqref{25}
\\
 &  MHz  &  m/sec &  m/sec & dB/cm & dB/cm  &
\\
\hline 
PMMA &   6 & 2756.4 & 1401.5 & 4.97 & 13.64 & \checkmark
\\
PMMA & 10 & 2760.5 &1404.8 & 7.69 & 23.99 & \checkmark
\\
PMMA & 18 & 2764.2 & 1405.1 & 12.68 & 37.21 & \checkmark
\\
PMMA & 20 & 2765.1 & 1405.7 & 12.64 & 44.28 & \checkmark
\\
PMMA & 30 & 2765.5 & 1406.1 & 19.64 & 63.94 & \checkmark
\\
%poly(4-methyl pentene-1) 
polymer \#1& 1.8 & 2180 & 1080 & 1.4& 6.7 & \checkmark
\\
polymer \#2  & 1.8 & 2040 & 830 & 1.8& 15 & \checkmark
%\\
%~~~~butadiene-styrene) &&&&&& 
\\
\hline \hline
\end{tabular}
\caption{ $f = \omega/2\pi$. Velocity and attenuation data for 
PMMA at 22.2$^\circ$C and  atmospheric pressure \cite{Asay1969}.  The data for the other polymers are from \cite{Hartmann1980}, where
polymer \#1 is poly(4-methyl pentene-1) and 
polymer \#2 is poly(acrylonitrile-butadiene-styrene).}
\label{table2}
\end{center}
\end{table}
We  consider the passivity constraint \eqref{25} in light of some reported ultrasonic data \cite{Asay1969,Hartmann1980}  in Table \ref{table2}.  
The values of absorption indicates loss moduli  of 1\% or less than the real parts, i.e. small attenuation for which the criterion \eqref{25}  applies, and is met for the data in Table \ref{table2}.   Velocity and attenuation data for 
styrene-butadiene rubber \cite{Wada1962} at 1 MHz over a temperature range from $0^\circ$ to $20^\circ$ is consistent with \eqref{25}.

Laymen et al.\ \cite{Layman2006} provide curve-fitted equations, eqs.\ (15-18) in \cite{Layman2006},  for all four of the parameters in \eqref{25} based on ultrasonic measurements on a particulate composite sample over a broad frequency range (2 to 10 MHz).  It may be easily verified that these wave speeds and attenuations  satisfy the condition \eqref{25} over the entire range of frequencies considered. 
Measurements of high frequency (25 to 65 MHz) velocities and attenuation in passive materials
for ultrasonic transducers at room temperature are given in \cite{Wang2001}.  The materials include
alumina/EPO-TEK 301 composites and tungsten/EPOTEK 301 composites.   We have checked that all of the parameters reported  satisfy eq.\ \eqref{25}.  
Pinton  et al.\ \cite{Pinton2012} measured attenuation and absorption of ultrasound in  skull bone.  They  reported longitudinal absorption of 2.7 dB/cm and  shear
absorption of 5.4 dB/cm at the  assumed longitudinal and shear wave speeds of 3,000 m/s and 1,500 ms, respectively, which clearly satisfies the positive absorption condition \eqref{25}.

\onecolumngrid %%%%%%%%%%%%%%%%%%%%%%%%%%%%%%%%%%%%%%%%%%%%%%%%%%%%%%%%%%%%%%%%%%

\begin{figure}[H]%htbp]
\centering
%\vspace{-1.2in}
%\hspace{-.1in}
\begin{subfigure}[b]{.5\textwidth}
\includegraphics[width=3.3in]{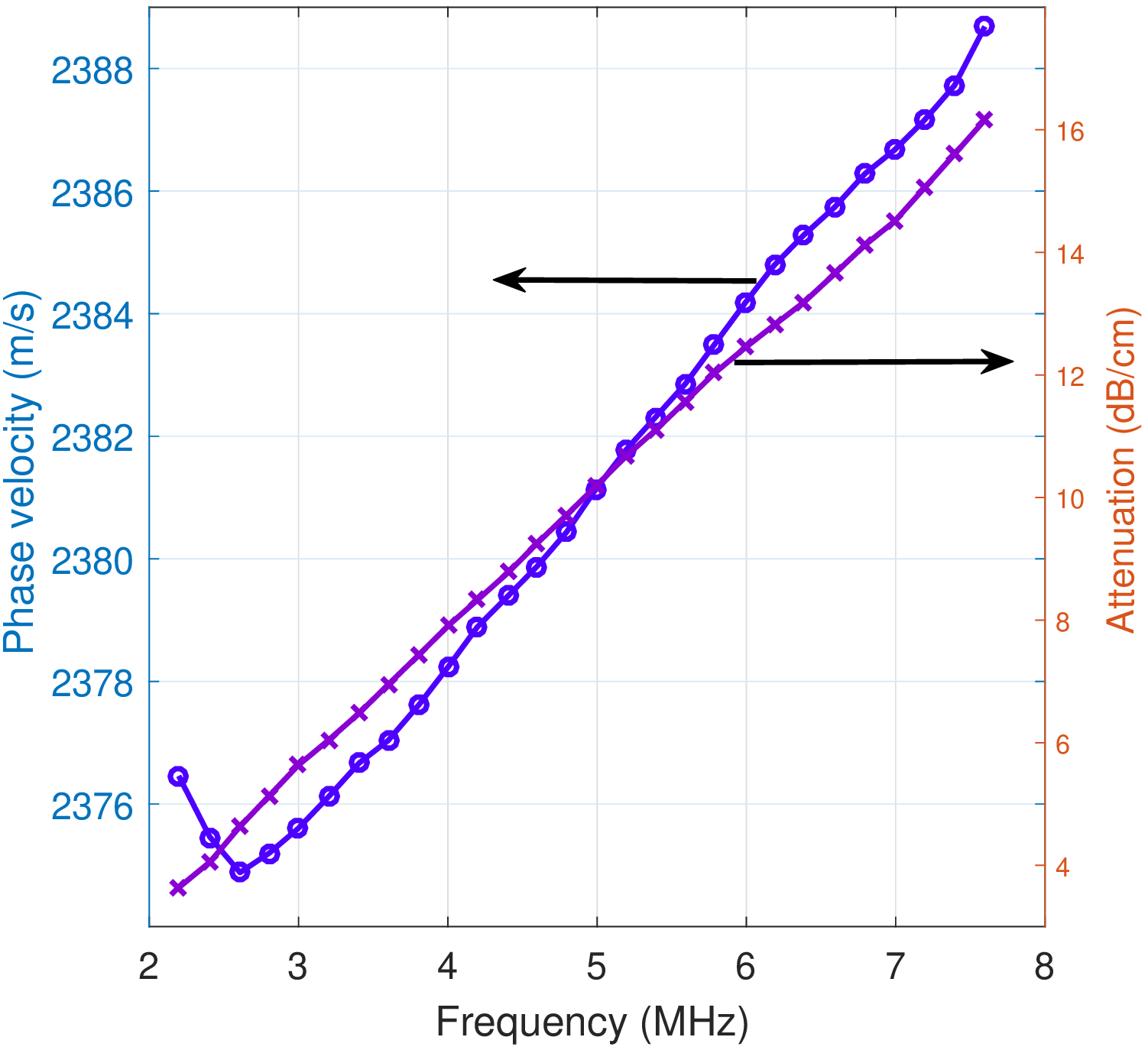}
\end{subfigure}
%\caption{Longitudinal}
\hspace{-.1in}
\begin{subfigure}[b]{.5\textwidth}
\includegraphics[width=3.3in]{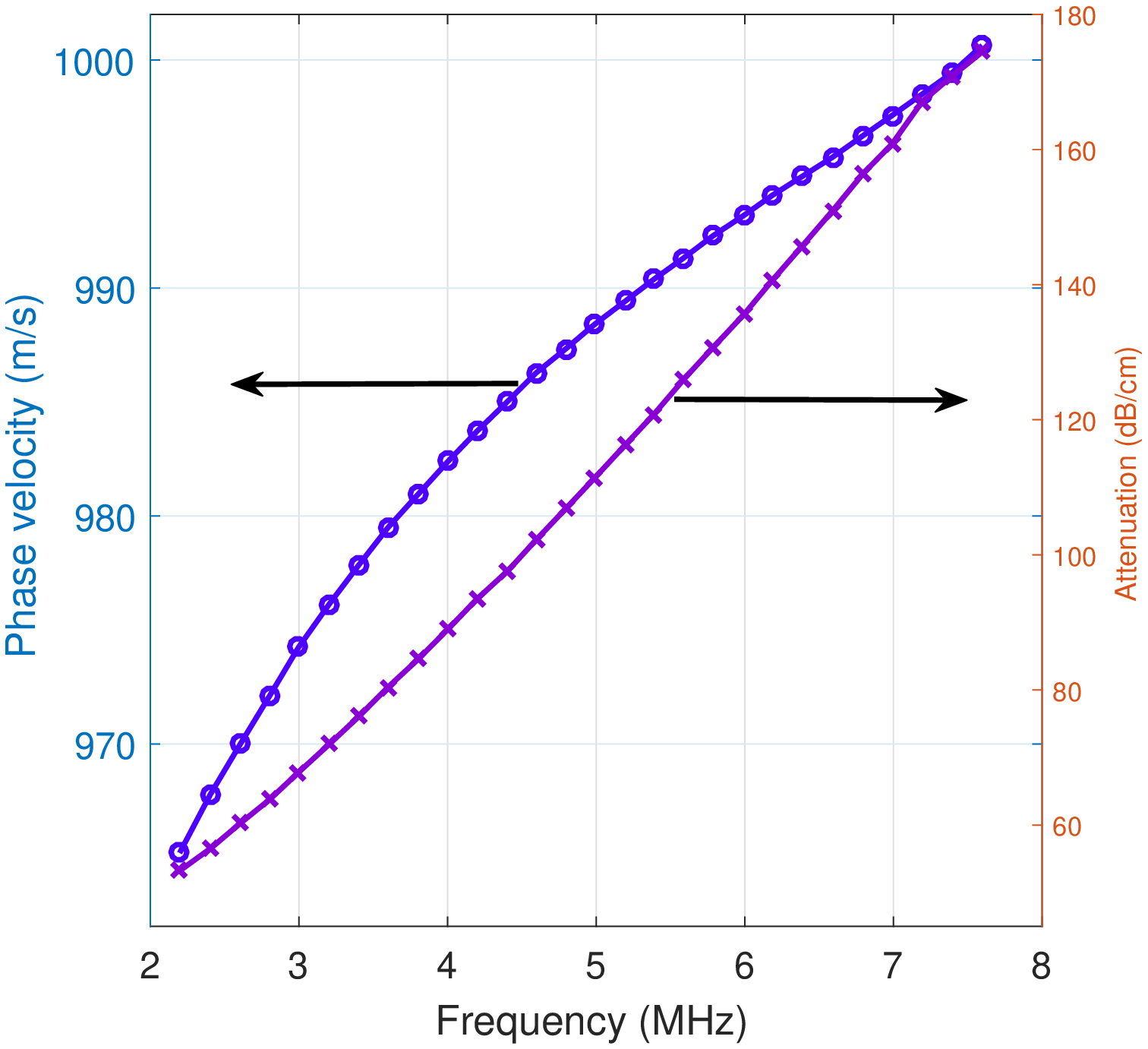}
\end{subfigure}
\\
%\vspace{-1.4in}
\caption{Phase velocity and attenuation as functions of frequency for  a high-density polyethylene sample: longitudinal (left) and transverse (right).  Data from \cite[Figure 4]{Wu1996}.   }
\label{fig1}
\end{figure}
\twocolumngrid%%%%%%%%%%%%%%%%%%%%%%%%%%%%%%%%%%%%%%%%%%%%%%%%%%%%%%%%%%%%%%%%%%

\begin{figure}[H]
\centering
%\vspace{-.85in}
%\hspace{-1.4in}
\includegraphics[width=3.6in]{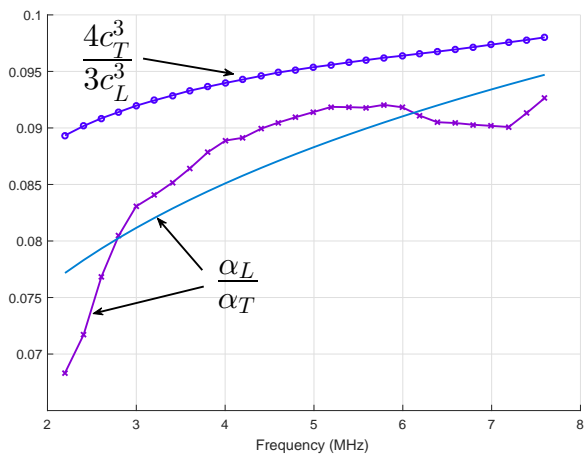}
%\caption{Longitudinal}
\\
%\vspace{-.8in}
\caption{The two terms in the passivity inequality \eqref{25}  calculated from the data of Figure \ref{fig1}.  The continuous curve for $\alpha_L/\alpha_T$ is based on eq.\ \eqref{-25} and Table \ref{table3}.}
\label{fig2}
\end{figure}
 
As a final example  we consider the velocity and attenuation data shown in Figure \ref{fig1}, which reproduces as accurately as possible the measurements reported in \cite{Wu1996} for a high-density polyethylene sample.  
The calculated values of the two terms in eq.\ \eqref{25} 
are plotted in Figure \ref{fig2}. 
The smooth curve for $\alpha_L/\alpha_T$ uses a fitted 
 power law model for attenuation proposed by Szabo and Wu \cite{Szabo2000}  and applied to the 
data of Wu \cite{Wu1996}.  The model assumes the attenuation has frequency dependence 
 of the form 
\beq{-25}
\alpha_M = \alpha_{0M} + \alpha_{1M} \, |f|^{y_\text{\tiny M}} , \ \ M = L,\, T, 
\eeq
where the coefficients for the high-density polyethylene sample are in Table \ref{table3}. 
Note that the formula corresponding to eq.\ (\ref{-25}) in Szabo and Wu \cite{Szabo2000} has $|\omega|$ instead of $|f|$, but we find that the numbers reported there are for eq.\ (\ref{-25}).
\begin{table}[h]
\begin{center}
\begin{tabular}{crcccccc} %{llllll}
\hline \hline
%\\
%Material &  
 $c_L$  &  $c_T$   & $\alpha_{0L}$ &  $\alpha_{1L}$  & $y_L$  & $\alpha_{0T}$ &  $\alpha_{1T}$  & $y_T$
\\
\hline 
%Lexan &   2195 & 943 & 0 & 4.783 & 1.001 & & 41.84 & 0.695
%\\
%Low density  &2566 &1273  & 0 & 7.577 & 0.796 & & 28.64 &0.815 
%\\
%polyethylene ~~ &&&&&&&&
%\\
%High density  & 
 2380 &987 & 0 &  1.522  &  1.171 & -0.517 & 22.80 & 1.00  %0.950
%\\ polyethylene &&&&&&&&
\\
\hline \hline
\end{tabular}
\caption{Absorption coefficients $\alpha_{0M}$ (dB/cm),  $\alpha_{1M}$ (dB/(MHz)$^{y_\text{\tiny M}}$ -cm), 
and exponential powers $y_\text{\tiny M}$ for high density polyethylene.    
Speeds, in m/s, are for the reference  frequency 4.8 MHz. Data from \cite{Szabo2000} except for $\alpha_{0T}$ which is not given there but is found here by least square fitting, and  $y_T$
which is found to provide better accuracy for $\alpha_T$ than the value 0.95 in \cite{Szabo2000}.}
\label{table3}
\end{center}
\end{table}

It is evident from the relative positions of the curves in Figure \ref{fig2} that the passivity inequality \eqref{25} is not satisfied 
%by the parameters of Figure \ref{fig1} 
at any of the frequencies considered.   We note that eq.\ \eqref{25} is an approximation valid for small values of attenuation.  The precise condition $\gamma \ge 0$ may be written in similar form as 
\beq{2=1}
\frac{\alpha_L}{\alpha_T} \ge 
\frac { 4c_{T}^3}  { 3c_{L}^3} \, I
\ \ \text{where} \ I = \Big( \frac{1+\gamma_L^2}{1+\gamma_T^2}\Big)^2 .
\eeq
Generally, the factor $I$ is close to but slightly less than unity, with 
$0.9964 \le I \le 0.9970$ for the data of Figure \ref{fig1}.  The effect of including this term in Figure \ref{fig2}
is almost imperceptible, i.e.,   the data is in violation of the  passivity condition for the bulk modulus. 
%The shear wave attenuation  is particularly high, with $0.347 \le \gamma_T \le 0.372$, in contrast to the relatively small longitudinal 
%wave attenuation  $0.063 \le \gamma_L \le 0.081$.   In view of the large shear attenuation the approximate conditions \eqref{231} and \eqref{25} are not valid.  In fact, we find that eq.\ \eqref{25} does not hold, see Figure \ref{fig2}. 
%However, as  Figure \ref{fig2} also shows, the correct inequality $\gamma >0$ is in fact satisfied at all frequencies, except for the lowest one.  The  
%This suggests that the data of  Figure \ref{fig1} is inconsistent in some manner.  
We can only conclude that the attenuation data for high-density polyethylene is not consistent with a passive linear viscoelastic model with frequency dependent complex-valued density and elastic moduli.

We note, however, that for the other data sets 
reported in \cite{Wu1996},   for samples of low-density polyethylene and Lexan Plexiglas,  we  find  the passivity condition \eqref{25} is satisfied.

% a0 = 0; a1= 1.522;  y =1.171 ;  a1= 22.8 ;  y =0.95;
% a = a0 + a1*(2*pi*f).^y

\section{Conclusion}
The main finding is  the constraint on the non-dimensional absorption parameter 
$\gamma$ in Eq.\ \eqref{23}.  For given values of wave speeds and shear absorption $ \gamma_T$ this sets a lower bound on the longitudinal absorption $ \gamma_L$.  The inequality $\gamma \ge 0$ has direct interpretation when absorption is small $(\gamma_L ,\, \gamma_T \ll 1)$, implying that the ratio of the attenuations per unit length, $\alpha_L /\alpha_T$, has a lower bound that depends on the ratio of the wave speeds, Eq.\ \eqref{25}.  The lower bound tends to zero as the Poisson's ratio of the material tends to $\frac 12$  $(\Leftrightarrow c_T/c_L \to 0)$. 
For instance, measurements of ultrasonic properties of soft tissues and tissue-like materials \cite{Madsen1983} shows shear wave attenuation coefficients on the order of $10^4$ times  the longitudinal wave attenuation coefficients.  Equation \eqref{25} then implies, under the small absorption assumption, that the transverse wave speed must be less than 4.2\% the value of the longitudinal speed.

%\acknowledgments Thanks to Philip L. Marston and Farid G. Mitri for interesting me in the topic, and to Michael A. Haberman for discussion. 

%%%%%%%%%%%%%%%%%%%%%%%%%%%%%%%%%%%%%%%%%%%%%%%%%%%%%%%%

%%%%%%%%%%%%%%%%%%%%%%%%%%%%%%%%%%%%%%%%%%%%%%%%%%%%%%%%%%%%%%%%%%%%%%%%%%
%\bibliography{../../SHARED_BIBLIOGRAPHY/AN_BIG_BIB}
%\bibliographystyle{unsrt}%natbib}%unsrtnat}%doipubmed}%harvard}% plain}%uabbrvnat}%

\end{document}